\documentclass[aps,prl,twocolumn,floatfix,superscriptaddress]{revtex4-2}

\usepackage{graphicx}
\usepackage[utf8]{inputenc}
\usepackage{amsmath,amsfonts}

\begin{document}

\author{M. Omidian}
\affiliation{Institut für Physik, Technische Universität Ilmenau, D-98693 Ilmenau, Germany}

\author{S. Leitherer}
\email{slei@nanotech.dtu.dk}
\affiliation{Center of Nanostructured Graphene, Department of Physics, Technical University of Denmark, DK-2800 Kongens Lyngby, Denmark}

\author{N. Néel}
\affiliation{Institut für Physik, Technische Universität Ilmenau, D-98693 Ilmenau, Germany}

\author{M. Brandbyge}
\affiliation{Center of Nanostructured Graphene, Department of Physics, Technical University of Denmark, DK-2800 Kongens Lyngby, Denmark}

\author{J. Kröger}
\email{joerg.kroeger@tu-ilmenau.de}
\affiliation{Institut für Physik, Technische Universität Ilmenau, D-98693 Ilmenau, Germany}

\title{Electric-field control of a single-atom polar bond}

\begin{abstract}
The polar covalent bond between a single Au atom terminating the apex of an atomic force microscope tip and a C atom of graphene on SiC(0001) is exposed to an external electric field.
For one field orientation the Au--C bond is strong enough to sustain the mechanical load of partially detached graphene, whilst for the opposite orientation the bond breaks easily.
Calculations based on density functional theory and nonequilibrium Green's function methods support the experimental observations by unveiling bond forces that reflect the polar character of the bond.  
Field-induced charge transfer between the atomic orbitals modifies the polarity of the different electronegative reaction partners and the Au--C bond strength.
\end{abstract}

\maketitle

Exploring the impact of electric fields on the modification of chemical-bond strengths is important for, e.\,g., electron transport across contacts in miniaturized devices and circuits \cite{natrevphys_1_211}.
Previously, the bond strength was controlled by modifying the number of covalent bonds atom by atom with the tip of a scanning tunneling microscope (STM) giving rise to characteristic changes in the overall junction conductance \cite{prl_104_176802,natnanotechnol_6_23}.
Electric-field effects and electrostatic forces have moreover been identified as notable ingredients for interpreting the contrast mechanism of an atomic force microscope (AFM) tip terminated by a single CO molecule \cite{science_325_1110,science_340_1434,science_348_308,jacs_137_7424,prl_113_226101,prl_115_076101,nl_16_1974}.

Given the relevance of individual chemical bonds and electrostatic effects in junctions at the ultimate scale it is desirable to explore the response of a single bond to an external electric field.
The combined experimental and theoretical work presented here provides direct evidence for the influence of an electric field on the strength of a polar covalent bond between two atoms.
An AFM is used to form and break in a controllable manner a bond between the Au atom terminating the AFM tip and a C atom of graphene on SiC(0001)\@.
The electric field is supplied by applying a voltage across the atomic junction.
An electric field pointing from the C to the Au atom (positive sample voltage $V_{\text{sample}}$) yields  a strong Au--C bond that enables the detachment of graphene from the surface in the course of tip retraction; the opposite field direction (negative $V_{\text{sample}}$), in contrast, induces a weak and easily breakable Au--C bond.
Density functional calculations including biased electrodes trace these observations to short-range bond forces that result from the polar covalent Au--C chemical bond whose strength is determined by the field-dependent charge allocation at the atoms.

\begin{figure*}
\includegraphics[width=0.8\linewidth]{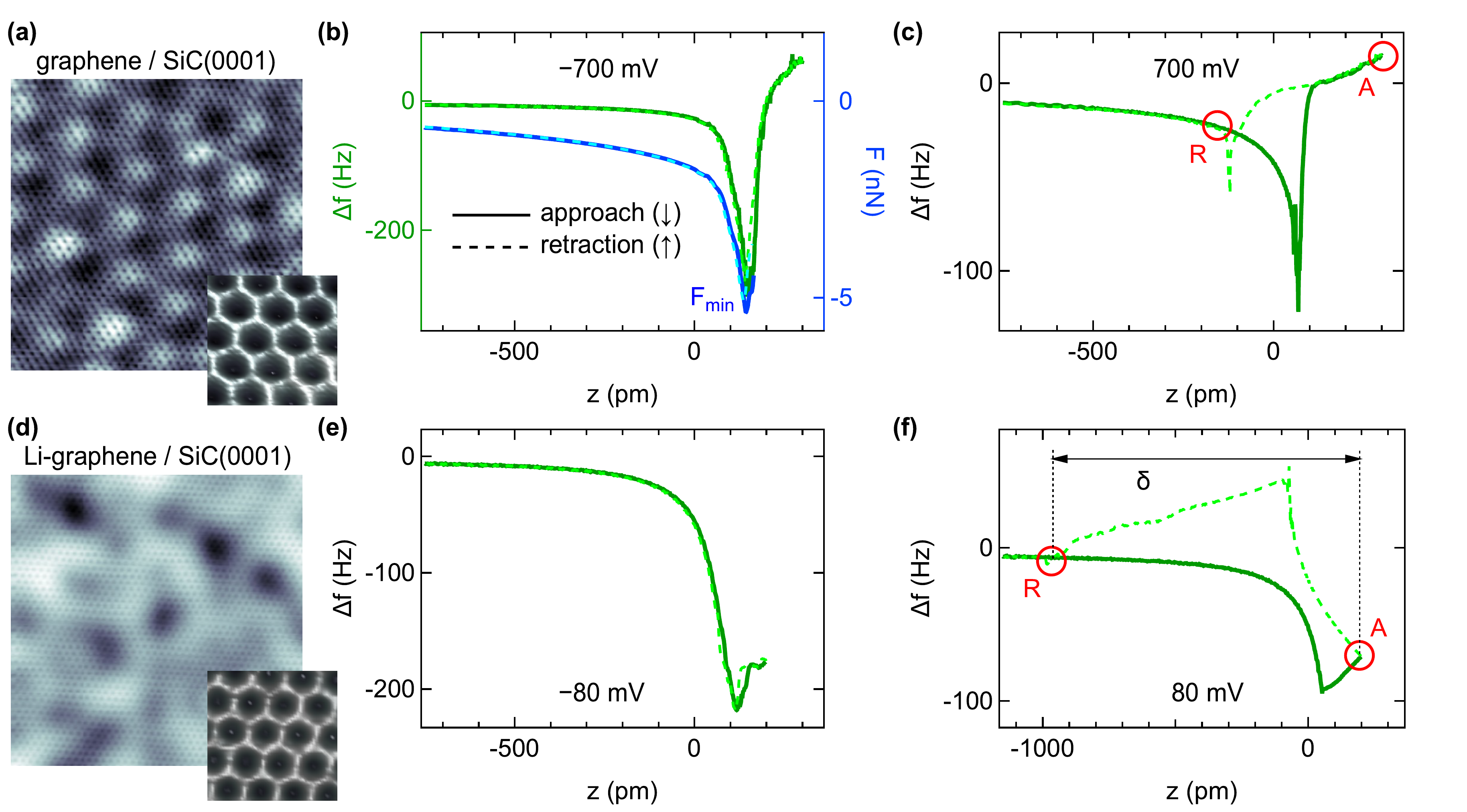}
\caption{
(a) Constant-current STM image of clean graphene [bias voltage: $-10\,\text{mV}$, tunneling current: $55\,\text{pA}$, size: $9.7\,\text{nm}\times 9.7\,\text{nm}$, the gray scale ranges from $0\,\text{pm}$ (black) to $67\,\text{pm}$ (white)] with close-up view of the graphene lattice [$1\,\text{mV}$, $33\,\text{pA}$, $0\,\text{pm}$ (black) -- $100\,\text{pm}$ (white)]\@.
(b) $\Delta f(z)$ for clean graphene, $V_{\text{sample}}=-700\,\text{mV}$ (approach: from left to right, upper solid line; retraction: from right to left, upper dashed line)\@.
The vertical force $F$ is displayed as the lower solid (approach) and dashed (retraction) line ($F_{\text{min}}$: point of maximum attraction)\@.
(c) $\Delta f(z)$ for clean graphene, $V_{\text{sample}}=700\,\text{mV}$\@ (\textit{A}: start of tip retraction; \textit{R}: intersection of $\Delta f_\downarrow$ and $\Delta f_\uparrow$ data.
(d) Like (a) for Li-intercalated graphene [$100\,\text{mV}$, $51\,\text{pA}$, $10\,\text{nm}\times 10\,\text{nm}$, $0\,\text{pm}$ (black) -- $81\,\text{pm}$ (white); close-up view: $100\,\text{mV}$, $50\,\text{pA}$, $0\,\text{pm}$ (black) -- $32\,\text{pm}$ (white)]\@. 
(e) $\Delta f(z)$ for Li-intercalated graphene, $V_{\text{sample}}=-80\,\text{mV}$\@.
(f) Like (e), $V_{\text{sample}}=80\,\text{mV}$\@.
The loop width of $\Delta f(z)$ is $\delta$.
In (b), (c) and (e), (f) $z=0\,\text{pm}$ is defined as the tip position prior to deactivating the feedback control and retracting the tip into the tunneling range ($z<0\,\text{pm}$)\@. 
}
\label{fig1}
\end{figure*}

Figure \ref{fig1} summarizes the first part of novel experimental results for clean [Fig.\,\ref{fig1}(a)--(c)] and Li-intercalated [Fig.\,\ref{fig1}(d)--(f)] graphene on SiC(0001) \cite{polbofo_sm_1}.
Clean graphene [Fig.\,\ref{fig1}(a)] yields STM images that are characterized by the previously reported $6\times 6$ superlattice with a spatial period of $1.72\pm 0.17\,\text{nm}$ \cite{surfsci_256_354,prl_100_176802} and the graphene lattice where the honeycomb cells are separated by $0.23\pm 0.02\,\text{nm}$, in agreement with expectations ($0.246\,\text{nm}$) \cite{njp_10_043033}.
For Li-intercalated graphene the $6\times 6$ superstructure is absent [Fig.\,\ref{fig1}(d)], signaling the efficient migration of Li through the Bernal-stacked graphene and C buffer layer on SiC(0001) \cite{prb_82_205402,prb_96_125429,surfsci_699_121638}.

Approaching the AFM tip towards the graphene lattice with $V_{\text{sample}}=-700\,\text{mV}$ and simultaneously recording the resonance frequency change, $\Delta f$, of the oscillating tuning fork leads to the distance-dependent data set referred to as $\Delta f_\downarrow(z)$ ($z$: tip displacement) in the following and depicted as the upper solid line in Fig.\,\ref{fig1}(b)\@.
The associated vertical force \cite{apl_78_123,apl_84_1801}, $F_\downarrow(z)$, appears as the lower solid line.
Care has been taken to show well-posed force data by appropriately adjusting the maximum probed distance range with respect to the position of inflection points in the extracted force \cite{natnanotechnol_13_1088}.
The minimum signals the point of maximum attraction.
Beyond contact the evolution of $F_\downarrow(z)$ deviates from the expected Lennard-Jones behavior, which would exhibit a steep increase due to Pauli repulsion.
Most likely, atomic relaxations of the junction geometry are the cause for the deviations.
Retraction of the AFM tip gives rise to $\Delta f_\uparrow(z)$ and $F_\uparrow(z)$ data that are depicted as dashed lines in Fig.\,\ref{fig1}(b)\@.
Obviously, approach and retraction data virtually coincide, that is, $\Delta f_\downarrow(z)\approx\Delta f_\uparrow(z)$ and $F_\downarrow(z)\approx F_\uparrow(z)$\@.
Using the opposite polarity of $V_{\text{sample}}$ [Fig.\,\ref{fig1}(c)] leads to a significantly different behavior of $\Delta f_\downarrow(z)$ and $\Delta f_\uparrow(z)$\@.
Rather than reaching a well defined minimum, $\Delta f_\downarrow$ abruptly changes its slope \cite{polbofo_sm_2}.
Commencing the tip retraction at point \textit{A}, $\Delta f_\uparrow(z)$ data do not reproduce $\Delta f_\downarrow(z)$ in the contact region.
The $\Delta f_\uparrow$ trace intersects $\Delta f_\downarrow$ at point \textit{R} before coinciding with $\Delta f_\downarrow(z)$ for further retraction giving rise to a $\Delta f$ loop.

A similar trend of $\Delta f_\downarrow$ and $\Delta f_\uparrow$ was observed for Li-intercalated graphene [Fig.\,\ref{fig1}(d)], i.\,e., $\Delta f_\downarrow(z)$ and $\Delta f_\uparrow(z)$ are essentially identical for $V_{\text{sample}}<0\,\text{V}$ [Fig.\,\ref{fig1}(e)] and strongly deviate from each other for $V_{\text{sample}}>0\,\text{V}$ [Fig.\,\ref{fig1}(f)]\@.
The loop width spanned by the distance between \textit{A} and \textit{R}, $\delta=z_A-z_R$, is, however, larger for Li-intercalated graphene than for its pristine counterpart\@.

The presented contact experiments are reproducible.
Subsequent approach-retraction cycles using the same tip, $V_{\text{sample}}$ and contact site yield virtually identical $\Delta f(z)$ data and leave the structural integrity of tip and sample invariant \cite{polbofo_sm_3}.
Therefore, forming and breaking the covalent tip--graphene bond is reversible.
Moreover, the current evolution $I(z)$ across the junction exhibits a consistent loop behavior \cite{polbofo_sm_2}.

Before discussing the $V_{\text{sample}}$ dependence of the $\Delta f$ loop, a tentative interpretation of the experimental observations shall be offered here and corroborated below by the simulations.
It seems that the chemical bond formed upon tip approach at $V_{\text{sample}}>0\,\text{V}$ is strong enough to locally detach the graphene sheet upon tip retraction.
Therefore, the $\Delta f_\downarrow$ data necessarily differ from $\Delta f_\uparrow$ for those distances where the graphene sheet is partially attached to the tip.
The point where the $\Delta f(z)$ loop closes [\textit{R} in Fig.\,\ref{fig1}(c),(f)] would then correspond to the release of the lifted graphene.
At $V_{\text{sample}}<0\,\text{V}$, in contrast, $\Delta f_\downarrow$ and $\Delta f_\uparrow$ nearly coincide, i.\,e., the chemical bond formed between the tip and graphene is weak and easily broken by tip retraction -- the graphene sheet remains on the surface and impedes the evolution of a $\Delta f$ loop.

\begin{figure}
\includegraphics[width=0.95\linewidth]{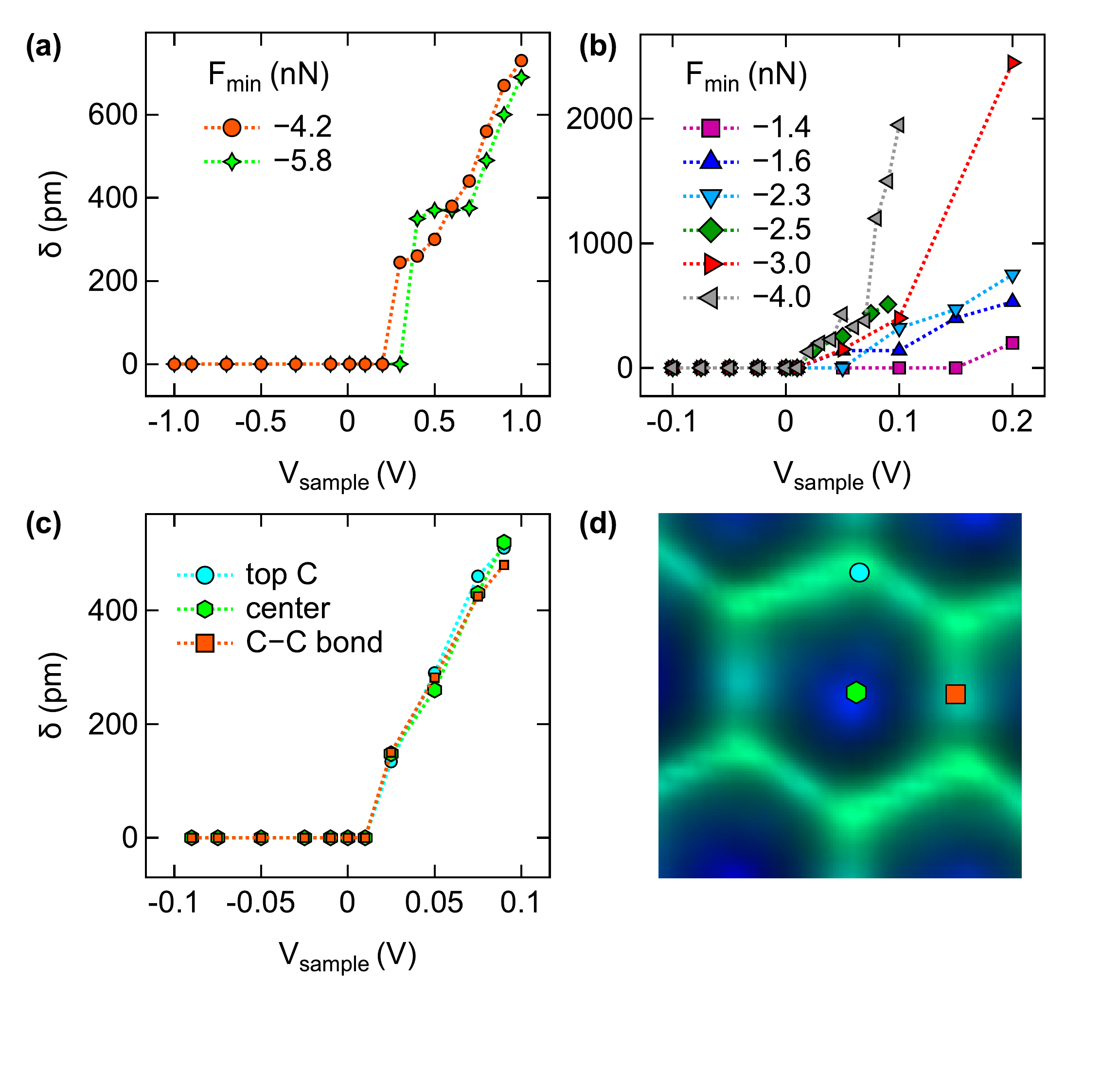}
\caption{
(a) $\delta(V_{\text{sample}})$ for clean graphene.
(b) Like (a) for Li-intercalated graphene.
Different tips in (a) and (b) are characterized by $F_{\text{min}}$\@.
(c) Site dependence of $\delta(V_{\text{sample}})$ for Li-intercalated graphene.
(d) Constant-height $\Delta f$ map [$0.47\,\text{nm}\times 0.47\,\text{nm}$, $-8.7\,\text{Hz}$ (dark) -- $7.5\,\text{Hz}$ (bright)] of a graphene honeycomb cell.
}
\label{fig2}
\end{figure}

To further characterize the $\Delta f$ behavior upon tip approach and retraction and its dependence on the $V_{\text{sample}}$ polarity, several other aspects were explored and are presented as the second part of novel experimental results in Fig.\,\ref{fig2}.
Figure \ref{fig2}(a),(b) compares $\delta(V_{\text{sample}})$ for clean [Fig.\,\ref{fig2}(a)] and Li-intercalated [Fig.\,\ref{fig2}(b)] graphene for a variety of tips.
The different tips are characterized by the magnitude of $F_{\text{min}}$\@.
Repeatedly performed field emission on and indentations into a Au substrate presumably cover the PtIr tip apex with Au and lead to different macroscopic tip shapes.
Therefore, the long-range van der Waals interaction between tip and surface is altered, which is reflected by $F_{\text{min}}$.
Both samples exhibit an asymmetric evolution of $\delta$ with the sign of $V_{\text{sample}}$\@.
Whilst $\delta$ vanishes for $V_{\text{sample}}\leq 0\,\text{V}$ it starts to increase monotonically for $V_{\text{sample}}>0\,\text{V}$\@.
Moreover, $\delta$ stays comparably low for clean graphene.
At $V_{\text{sample}}=1\,\text{V}$, $\delta$ is still lower than $750\,\text{pm}$ [Fig.\,\ref{fig2}(a)]\@.
For Li-intercalated graphene and a tip with similar force minimum $F_{\text{min}}$ as in the case of clean graphene, $\delta$ adopts nearly $2000\,\text{pm}$ already at $V_{\text{sample}}=0.1\,\text{V}$ [Fig.\,\ref{fig2}(b)]\@.
This effect is most likely caused by graphene being in its quasi-free state, which facilitates its detachment from the surface.

For the Li-intercalated sample, $\delta(V_{\text{sample}})$ data obtained are collected in Fig.\,\ref{fig2}(b)\@.
A clear trend is visible.
An increase of $\vert F_{\text{min}}\vert$ entails a larger slope of $\delta$ for $V_{\text{sample}}>0\,\text{V}$, i.\,e., $\delta$ adopts large values already at low $V_{\text{sample}}$.
This observation is plausible because the long-range van der Waals force acts as an additional background attraction and assists in lifting the graphene sheet \cite{nl_10_461,science_336_1557,prb_91_195436}.
The strength of the covalent tip--graphene bond, however, is determined by the short-range bond force, as clarified by the calculations below.

In a different set of experiments the graphene lattice site dependence of the characteristic $\Delta f$ behavior was explored.
To this end, the high symmetry points of the honeycomb cell [a C atom, a C--C bond and a honeycomb center, Fig.\,\ref{fig2}(d)] were scrutinized.
Despite the different lattice sites probed, $\delta(V_{\text{sample}})$ is similar [Fig.\,\ref{fig2}(c)]\@.
This observation indicates a preferred bond configuration that is achieved by relaxations of atom positions both at the tip apex and the graphene lattice and that is, therefore, a configuration adopted independent of the approach position.  
Selectively enhanced chemical reactivity of graphene C atoms on a metal surface were inferred previously from tunneling-to-contact transitions in STM junctions \cite{prl_105_236101}.

\begin{figure}
\includegraphics[width=0.95\linewidth]{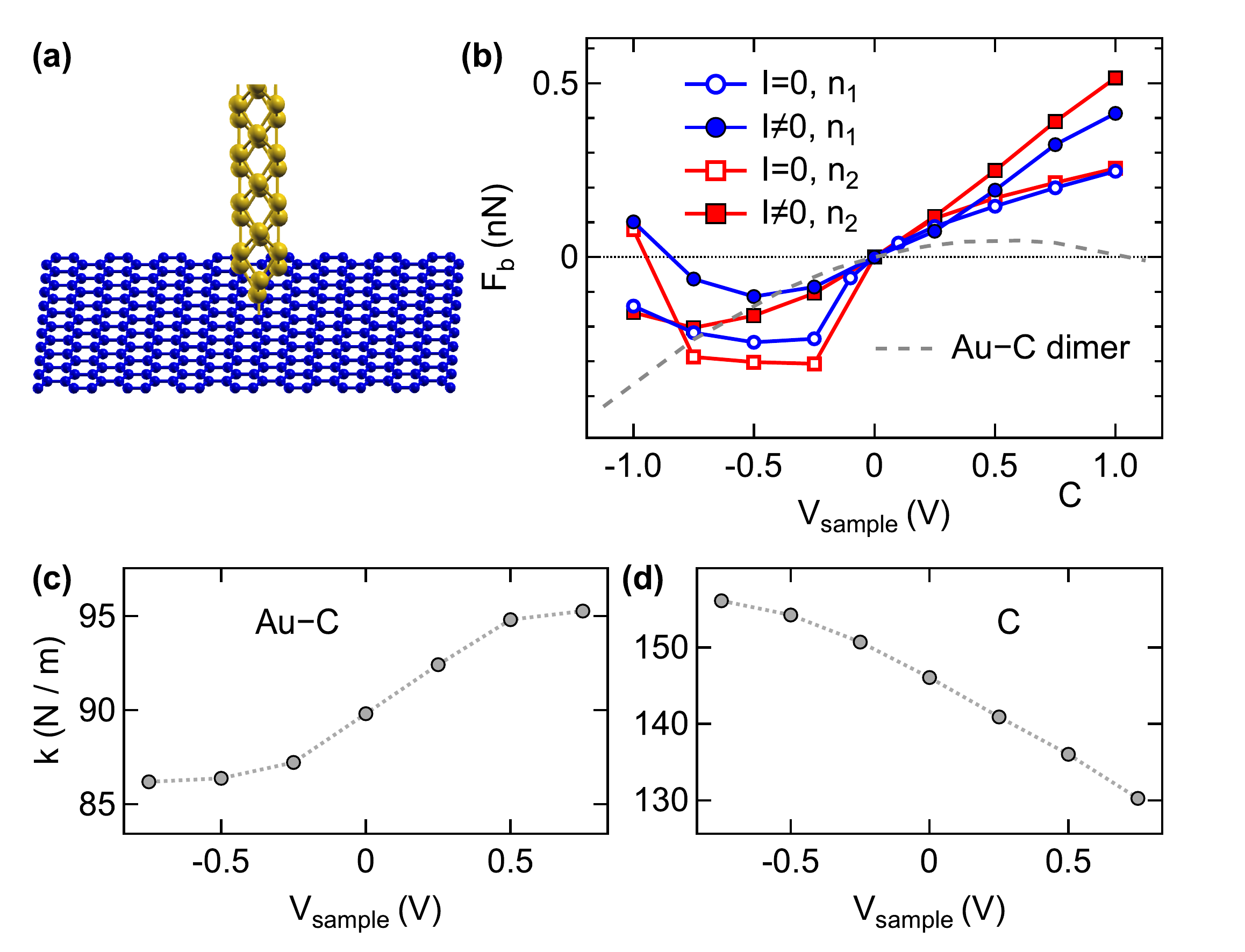}
\caption{
(a) Model geometry comprising a quasi-one-dimensional Au tip and free-standing graphene. 
(b) Calculated bond force $F_\text{b}(V_{\text{sample}})$ [$F_\text{b}(0\,\text{V})$ set to $0\,\text{nN}$] for the model geometry in (a) (solid lines) and a Au--C dimer (dashed line)\@. 
The electric field is applied perpendicular to the graphene plane.
Electron densities $n_1<n_2$ are adjusted by the Dirac point energy, $-400\,\text{meV}$ for $n_1$ and $-700\,\text{meV}$ for $n_2$.
Solid (Open) symbols reflect the presence (absence) of a current $I$\@.
(c) Spring constant $k_{\text{Au}-\text{C}}$ as a function of $V_{\text{sample}}$ for doping $n_2$. 
(d) Spring constant $k_\text{C}$ as a function of $V_{\text{sample}}$ calculated for displacing a C atom at doping $n_2$.
}
\label{fig3}
\end{figure}

The density functional and transport calculations \cite{polbofo_sm_4} were carried out for a simplified quasi-one-dimensional Au tip on top of a free finite graphene sheet [Fig.\,\ref{fig3}(a)] \cite{nanoscale_11_6153,prb_100_195417}.
They showed that the relaxation of the junction geometry prefers the top-C position to the hollow (by $\approx 0.2\,\text{eV}$) and bridge (by $\approx 0.04\,\text{eV}$) site of the graphene honeycomb cell.
Consequently, independent of the tip position atop the graphene honeycomb cell a bond configuration in which the tip-terminating Au atom is positioned atop a C atom is preferred.
This result is consistent with the experimental finding of a site-independent variation of $\delta$ with $V_{\text{sample}}$ [Fig.\,\ref{fig2}(c)]\@.

The atomic force induced by the applied field and the flowing current is calculated in the Born-Oppenheimer approximation and referred to as the bond force, $F_\text{b}$, in the following.
It is defined as the projection of $\mathbf{F}_{\text{Au}}-\mathbf{F}_{\text{C}}$ onto $\mathbf{r}_{\text{Au}-\text{C}}$, i.\,e., $F_\text{b}=\left(\mathbf{F}_{\text{Au}}-\mathbf{F}_\text{C}\right)\cdot\mathbf{r}_{\text{Au}-\text{C}}/\vert\mathbf{r}_{\text{Au}-\text{C}}\vert$, with $\mathbf{F}_{\text{Au}}$ ($\mathbf{F}_\text{C}$) the total force acting on the Au (C) atom and $\mathbf{r}_{\text{Au}-\text{C}}=\mathbf{r}_\text{C}-\mathbf{r}_{\text{Au}}$ the vector from Au to C\@.
The sign of $F_\text{b}$ is thus defined positive (negative) for attraction (repulsion)\@. 

In addition to the model setup [Fig.\,\ref{fig3}(a)] a Au--C dimer was considered in the calculations.
Both configurations reveal an unambiguous asymmetry of $F_\text{b}$ with $V_{\text{sample}}$ [Fig.\,\ref{fig3}(b)] -- it is repulsive for $V_{\text{sample}}<0\,\text{V}$ and attractive for $V_{\text{sample}}>0\,\text{V}$\@.
This behavior applies to different electron doping levels $n_1<n_2$ as well as to the presence or absence of a current $I$ across the junction.
In particular, all models reveal similar $F_\text{b}$ magnitudes and a bond strengthening for positive $V_{\text{sample}}$, while deviations arise mainly due to the different system sizes.
For $V_{\text{sample}}>0\,\text{V}$ the calculated data reveal a slightly larger $F_\text{b}$ for $n_2$ than for $n_1$\@.
The higher $n$-doping yields a stronger screening in graphene which tends to increase the voltage drop and local electric field at the Au--C contact.
This calculated result is compatible with the experimental observation of a more pronounced $\Delta f$ loop for Li-intercalated graphene. 

The sample voltage asymmetry of $F_\text{b}$ entails a corresponding asymmetry of the Au--C bond strength, which is plotted as the Au--C spring constant, $k_{\text{Au}-\text{C}}$, in Fig.\,\ref{fig3}(c)\@.
To obtain $k_{\text{Au}-\text{C}}$, the tip was displaced at constant $V_{\text{sample}}$ and the change in $F_\text{b}$ evaluated.
The increased $k_{\text{Au}-\text{C}}$ at $V_{\text{sample}}>0\,\text{V}$ supports the idea of graphene detachment upon tip retraction.
The detachment scenario is further corroborated by the evolution of the spring constant of a graphene C atom, $k_\text{C}$, with $V_{\text{sample}}$ [Fig.\,\ref{fig3}(d)]\@, which was obtained by the finite displacement of the C atom along the surface normal. 
For increasing $V_{\text{sample}}$, $k_\text{C}$ becomes significantly smaller.
Therefore, the C atom is more easily moved due to the attraction to the Au atom for $V_{\text{sample}}>0\,\text{V}$ than for $V_{\text{sample}}<0\,\text{V}$.

\begin{figure}
\includegraphics[width=0.95\linewidth]{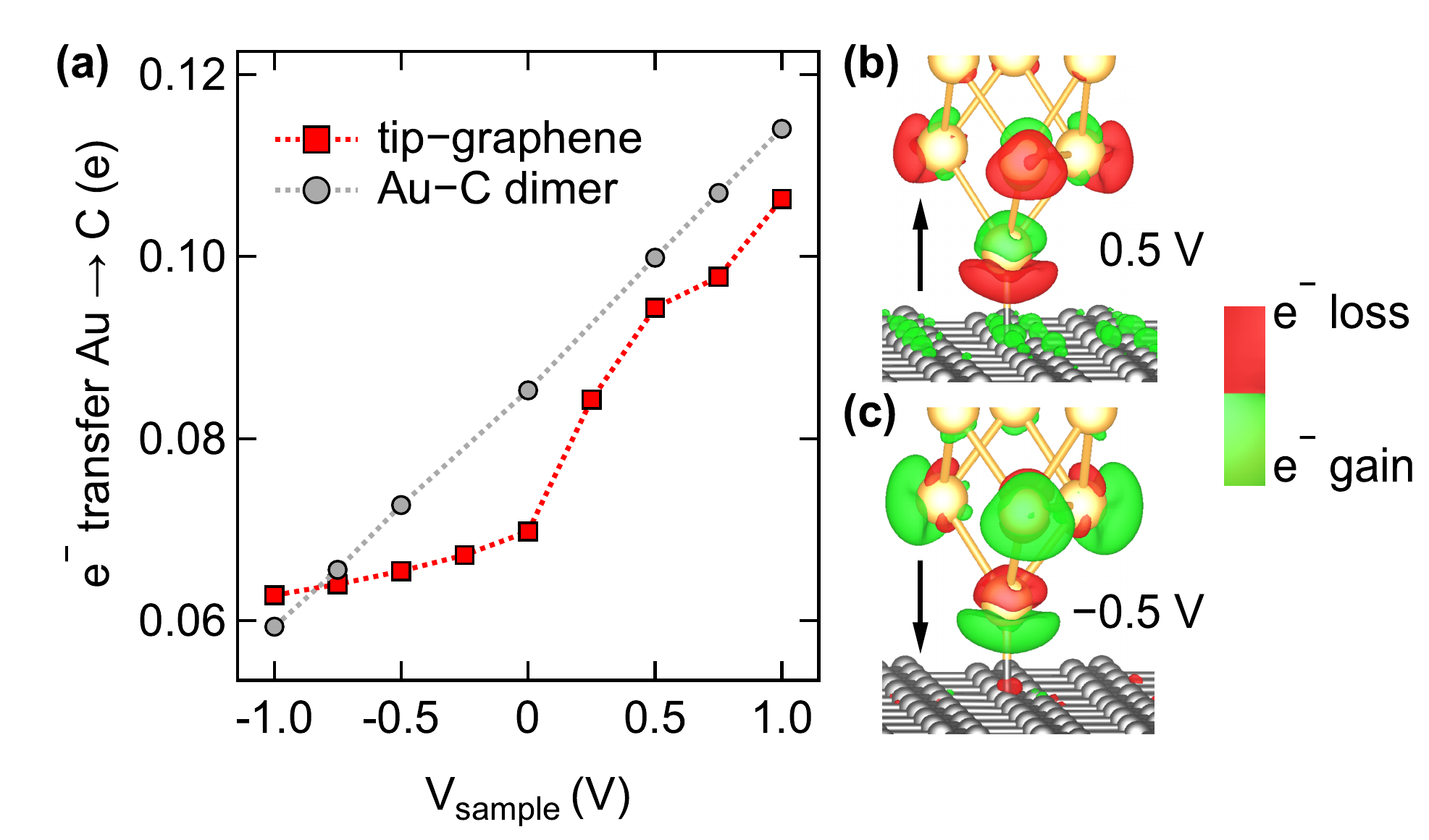}
\caption{
(a) Hirshfeld charge analysis of electron transfer from Au to C depending on $V_{\text{sample}}$ calculated for the tip--graphene model (squares) and the Au--C dimer (dots)\@. 
(b) Induced electron density with respect to $0\,\text{V}$ for the tip--graphene model at $V_{\text{sample}}=0.5\,\text{V}$\@.
Red (Green) shaded areas depict electron loss (gain) with a density isosurface of $20\,\text{e}/\text{nm}^3$ (e: elementary charge)\@.
(c) Like (b), $V_{\text{sample}}=-0.5\,\text{V}$\@.
The arrow in (b) and (c) shows the electric-field direction.
}
\label{fig4}
\end{figure}

An important question to be answered concerns the origin of the $V_{\text{sample}}$ asymmetry of $F_\text{b}$.
The charge transfer in the Au--C dimer [dots in Fig.\,\ref{fig4}(a)] as well as between the Au tip and graphene (squares) was calculated in a Hirshfeld charge analysis \cite{tca_44_129}. 
At $V_{\text{sample}}=0\,\text{V}$, i.\,e., at zero electric field, electrons are transferred from (less electronegative) Au to (more electronegative) C of the dimer leading to an electric dipole \cite{jacs_141_342}.
For $V_{\text{sample}}>0\,\text{V}$ ($V_{\text{sample}}<0\,\text{V}$) electron transfer from Au to C is enhanced (reduced) compared to $0\,\text{V}$\@.
For the tip--graphene model the electron transfer from the Au tip to graphene follows the same trend, with $V_{\text{sample}}>0\,\text{V}$ supporting the electron transfer from the Au tip to graphene, which accumulates positive charge at the tip apex and negative charge in the atomic environment of the contacted graphene C atom [Fig.\,\ref{fig4}(b)]\@.
The zero-bias dipole is enhanced and the polar bond is strengthened for $V_{\text{sample}}>0\,\text{V}$\@.
For $V_{\text{sample}}<0\,\text{V}$, in contrast, electron transfer from Au to C is hindered and accumulates the opposite charge at the atoms [Fig.\,\ref{fig4}(c)]\@. 
The polar bond is thus weakened.
This field-driven -- rather than current-induced -- effect is consistent with the experimentally observed nearly point-symmetric current with respect to $V_{\text{sample}}=0\,\text{V}$  in a $V_{\text{sample}}$ interval where the asymmetry of $\delta$ is already clearly visible \cite{polbofo_sm_5}.
Moreover, the calculated projected densities of states involving the relevant $p_z$, $d_{xz}$, $d_{yz}$ orbitals \cite{polbofo_sm_6} clearly reveal the field-induced charge redistribution and show that positive (negative) $V_{\text{sample}}$ tends to decrease (increase) the Au--C bond length \cite{polbofo_sm_7}.
This mechanism differs from the current-induced forces reported for homonuclear bonds \cite{prb_67_193104,prb_100_035415,nl_19_7845,smallmeth_4_1900817} where changes in bond strength were attributed to modifications in the overlap population caused by the nonequilibrium filling of scattering states, rather than to charge transfer and dipole field interaction between the components.

In conclusion, the combination of atomic force microscopy, density functional and nonequilibrium Green's function calculations unveils that the archteypical polar bond between two atoms can individually be influenced by the magnitude and orientation of an external electric field.
The control of a single polar chemical two-atom bond proceeds via the field-induced charge transfer between the atoms with different electronegativity.
Tailoring chemical bond strengths at the single-atom level together with the possibility of applying and releasing mechanical load open the path to locally distort matter and explore its response.
From a chemical point of view, reactivity and catalytic activity may be accessed at the atomic scale with the presented methods.

\acknowledgments{Funding by the Deutsche Forschungsgemeinschaft (Grant No.\ KR 2912/10-3) and Villum Fonden (Grant No.\ 00013340)\@ is acknowledged.
The Center for Nanostructured Graphene (CNG) is sponsored by the Danish Research Foundation (Grant No.\ DNRF103)\@.}

\bibliographystyle{apsrev4-2}
%

\end{document}


\author{M. Omidian}
\affiliation{Institut für Physik, Technische Universität Ilmenau, D-98693 Ilmenau, Germany}

\author{S. Leitherer}
\affiliation{Center of Nanostructured Graphene, Department of Physics, Technical University of Denmark, DK-2800 Kongens Lyngby, Denmark}

\author{N. Néel}
\affiliation{Institut für Physik, Technische Universität Ilmenau, D-98693 Ilmenau, Germany}

\author{M. Brandbyge}
\affiliation{Center of Nanostructured Graphene, Department of Physics, Technical University of Denmark, DK-2800 Kongens Lyngby, Denmark}

\author{J. Kröger}
\affiliation{Institut für Physik, Technische Universität Ilmenau, D-98693 Ilmenau, Germany}

\title{\normalfont{Supplemental Material for:} \\ Electric-field control of a single-atom polar bond}

\maketitle

\section{Experimental methods}

Experiments were performed with a combined AFM-STM operated in ultrahigh vacuum ($10^{-9}\,\text{Pa}$) and at low temperature ($5\,\text{K}$, $78\,\text{K}$)\@.
PtIr tips were attached to the free prong of a quartz tuning fork \cite{apl_73_3956} oscillating with a resonance frequency of $29.9\,\text{kHz}$, a quality factor of $45000$ and an amplitude of $50\,\text{pm}$.
Changes in the resonance frequency, $\Delta f$, were measured in noncontact frequency modulation experiments. 
The tips were prepared by field emission on and indentations into clean Au(111), which presumably led to a coating of the tip apex with several atomic layers of the substrate material.
The tip apex geometry was further trained by the transfer of single atoms from the tip to the surface ensuring its termination with a single atom \cite{prl_94_126102,jpcm_20_223001,pccp_12_1022}.
The \textit{n}-doped 6H-SiC(0001) crystal with epitaxially grown graphene residing atop a C buffer layer in the Bernal configuration was cleaned by annealing ($870\,\text{K}$)\@.
Lithium was evaporated from a resistively heated dispenser and intercalated at $570\,\text{K}$\@.
STM images of the sample surfaces were acquired in the constant-current mode with the bias voltage $V_{\text{sample}}$ applied to the sample.
AFM images represent $\Delta f$ maps acquired at constant height.
Images were processed using WSxM \cite{rsi_78_013705}.

\section{Vertical-distance evolution of the current}

\begin{figure}
\includegraphics[width=0.65\linewidth]{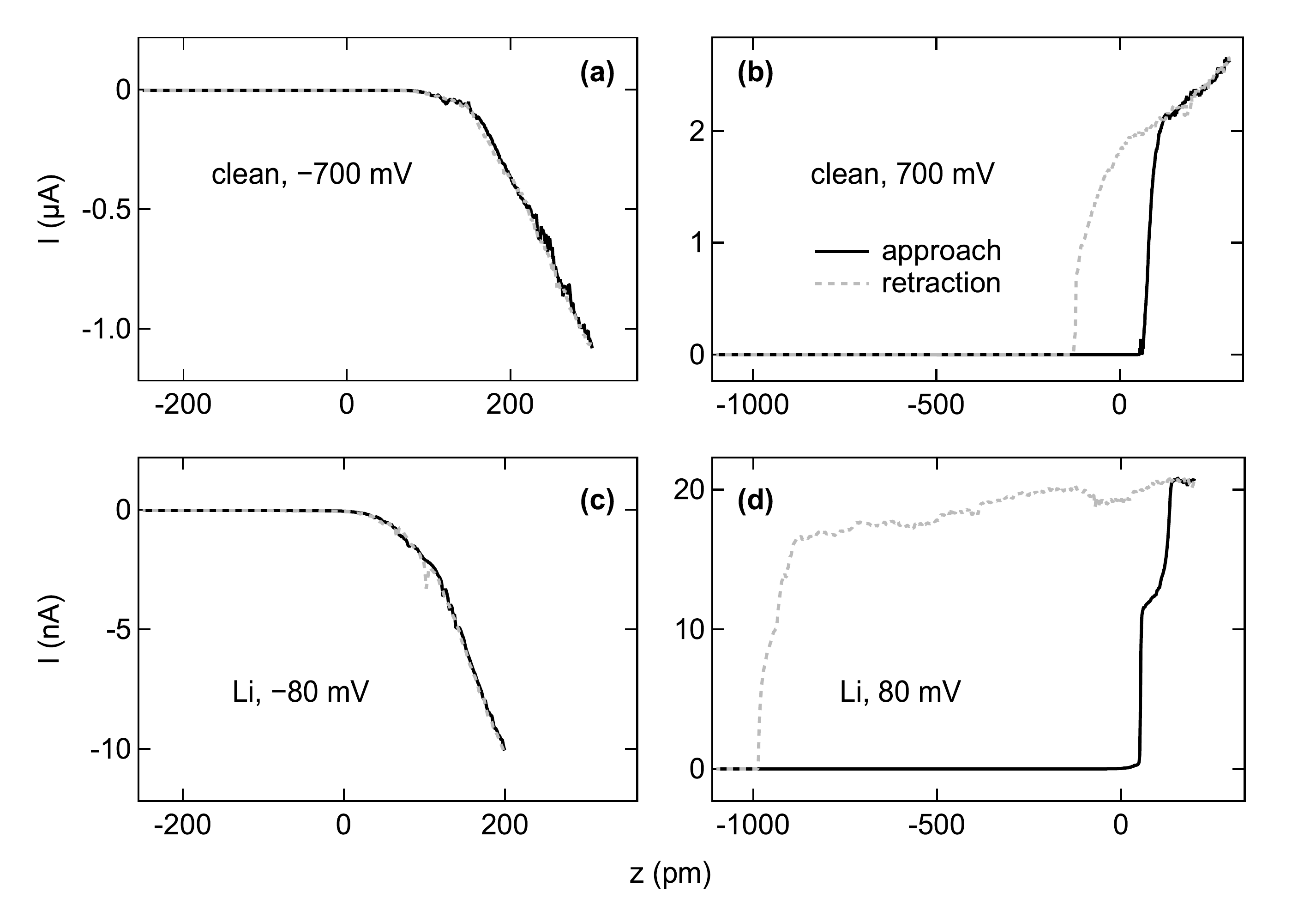}
\caption{
Evolution of the current $I$ with tip displacement $z$ for (a), (b) clean and (c), (d) Li-intercalated graphene on SiC(0001) for indicated $V_{\text{sample}}$\@.
Zero tip displacement is defined as the tip position prior to disabling the feedback loop and retracting the tip into the tunneling range ($z<0\,\text{pm}$)\@.
}
\label{figS1}
\end{figure}

Simultaneously with the resonance frequency change $\Delta f(z)$ the current evolution $I(z)$ was acquired.
Figure \ref{figS1} shows that $I(z)$ exhibits the same loop behavior as $\Delta f(z)$ (Fig.\,1 of the main article)\@.
In particular, at negative $V_{\text{sample}}$, approach [solid lines in Fig.\,\ref{figS1}(a),(c)] and retraction (dashed lines) currents are virtually identical, whilst positive $V_{\text{sample}}$ [Fig.\,\ref{figS1}(b),(d)] leads to a pronounced loop behavior where the retraction currents clearly deviate from the approach currents.

\begin{figure}[h]
\includegraphics[width=0.65\linewidth]{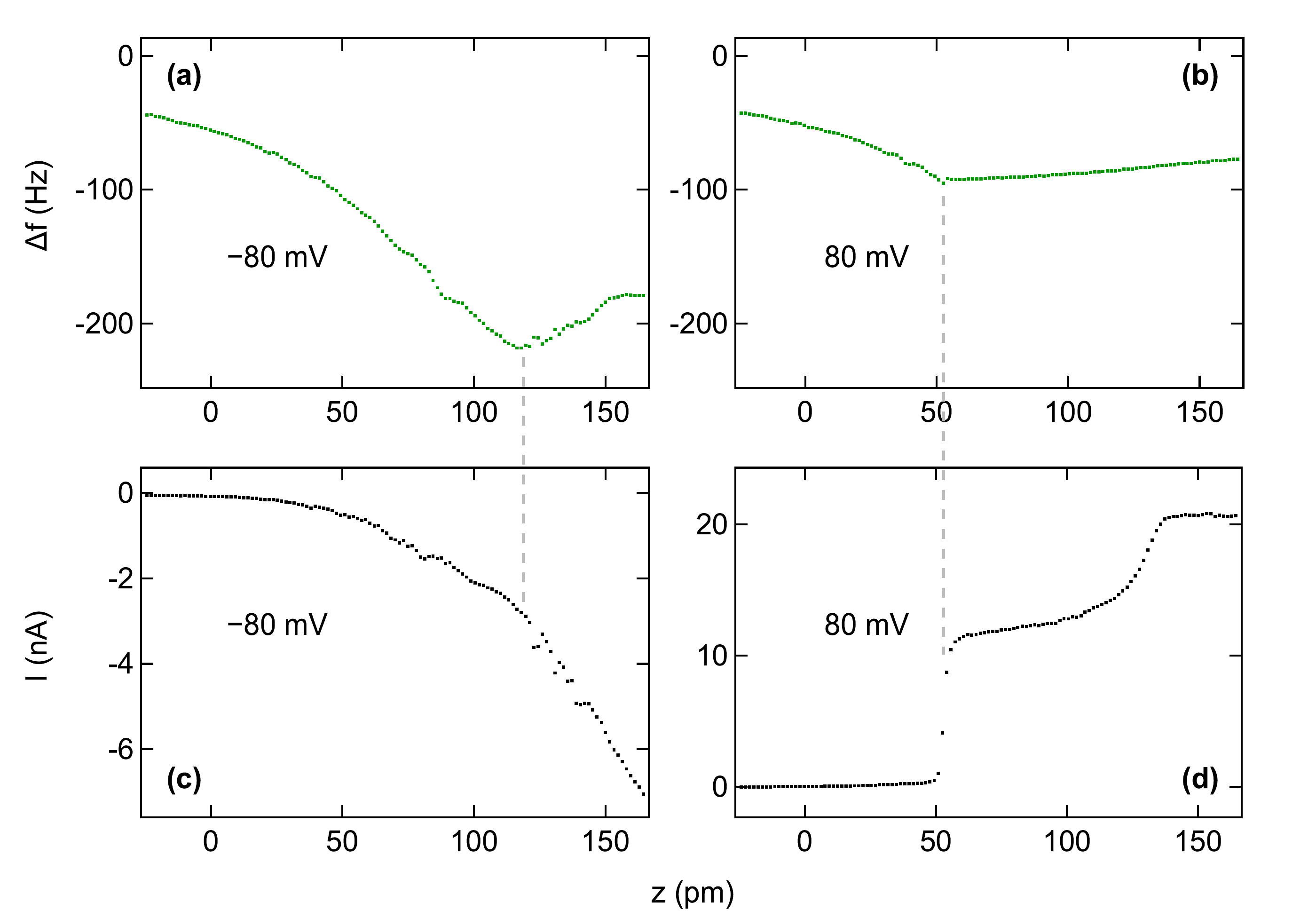}
\caption{
Close-up view of (a), (b) $\Delta f(z)$ and (c), (d) $I(z)$ in the vicinity of the point of maximum attraction for Li-intercalated graphene and indicated $V_{\text{sample}}$\@.
The dashed line in (a), (c) relates the $\Delta f(z)$ minimum at $z\approx 118\,\text{pm}$ with the gradual decrease of $I(z)$\@.
In (b), (d) the dashed line shows that the sudden change in the $\Delta f(z)$ slope at $z\approx 52\,\text{pm}$ is accompanied by an abrupt jump of $I(z)$ by nearly an order of magnitude.
}
\label{figS2}
\end{figure}

It is interesting to directly compare $\Delta f(z)$ and $I(z)$ in the vicinity of the point of maximum attraction.
Figure \ref{figS2} shows a representative example for Li-intercalated graphene.
For $V_{\text{sample}}=-80\,\text{mV}$, $\Delta f(z)$ exhibits a clearly developed local minimum at $z\approx 118\,\text{pm}$ [Fig.\,\ref{figS2}(a)], whilst at $80\,\text{mV}$ $\Delta f(z)$ abruptly changes its slope at $z\approx 52\,\text{pm}$, i.\,e., above the tip position where $\Delta f(z)$ reaches its minimum at $-80\,\text{mV}$\@.
At $V_{\text{sample}}=-80\,\text{mV}$, $I(z)$ gradually decreases without showing a fingerprint of the $\Delta f(z)$ minimum [dashed line connecting Fig.\,\ref{figS2}(a) and Fig.\,\ref{figS2}(c)]\@.
The situation is markedly different for $V_{\text{sample}}=80\,\text{mV}$ [Fig.\,\ref{figS2}(d)]\@.
At the tip position where $\Delta f(z)$ abruptly changes its slope, $I(z)$ increases in a nearly discontinuous manner.
The sudden jump of $I(z)$ is indicative of relaxations of the atomic-scale junction geometry \cite{prl_94_126102,jpcm_20_223001,prl_102_086805,pccp_12_1022,nl_11_3593}.
Another strong increase of $I(z)$ occurs at $z\approx 125\,\text{pm}$, which is accompanied by a gradual variation of $\Delta f(z)$\@.  

\section{Reversibility of bond formation}

\begin{figure}[h]
\includegraphics[width=0.65\linewidth]{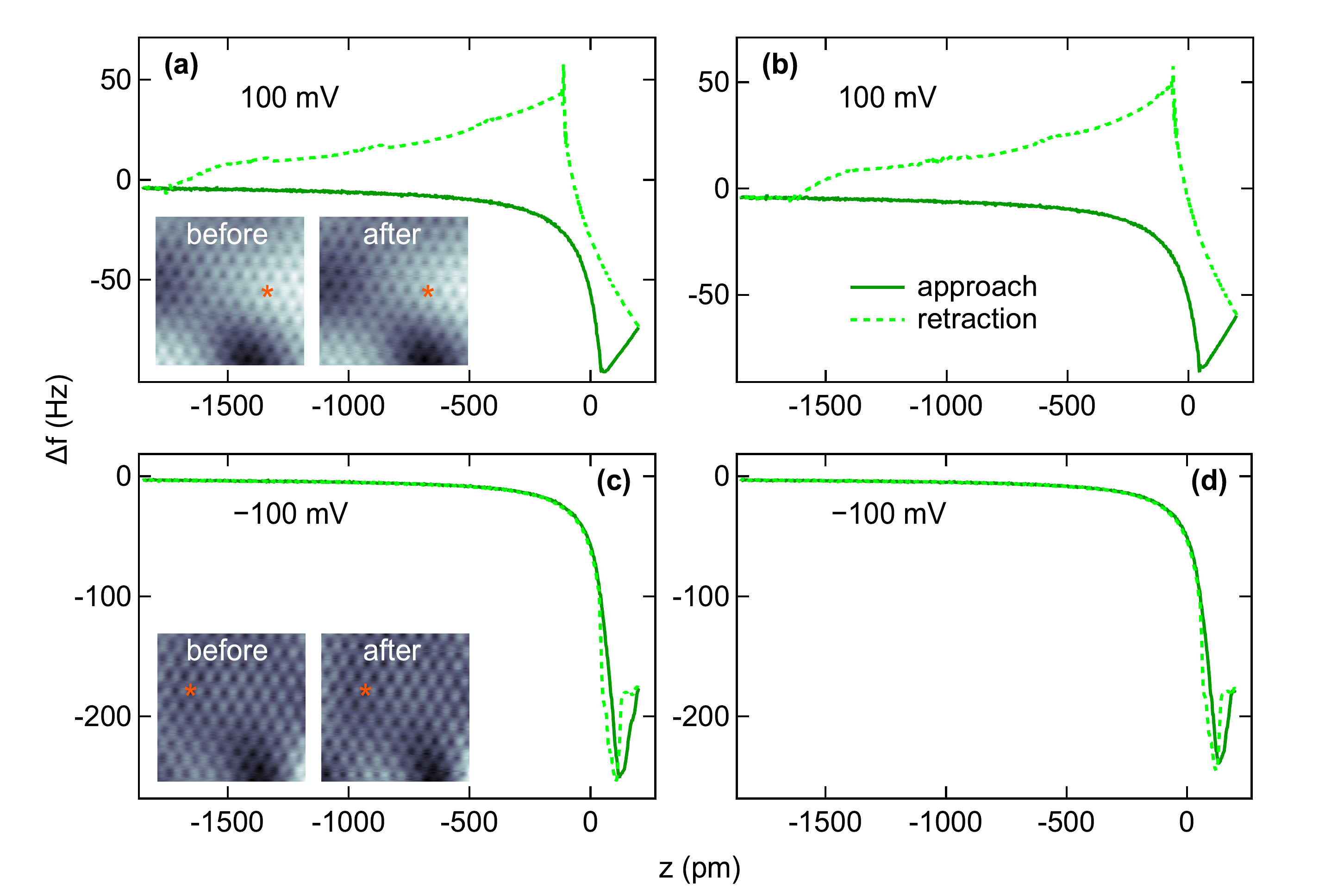}
\caption{
Subsequent approach-retraction cycles on Li-intercalated graphene performed at (a), (b) $V_{\text{sample}}=100\,\text{mV}$ and (c), (d) $V_{\text{sample}}=-100\,\text{mV}$\@.
In each panel the resonance frequency change $\Delta f$ is plotted as a function of the tip displacement $z$ ($z=0\,\text{pm}$ is defined as the tip displacement prior to disabling the feedback loop and tip retraction to a position in the tunneling range, $z<0\,\text{pm}$)\@.
Solid (Dashed) lines depict approach (retraction) data.
Insets to (a) and (c): STM images of graphene before and after a contact experiment with indicated contact position (asterisk) ($100\,\text{mV}$, $50\,\text{pA}$, $2.3\,\text{nm}\times 2.3\,\text{nm}$)\@.
}
\label{figS3}
\end{figure}

Using Li-intercalated graphene as a representative example, Fig.\,\ref{figS3} shows that the contact experiments presented in the main article are reproducible.
Surface and tip retain their structural integrity and the bond forming/breaking process is reversible.
To this end, $\Delta f(z)$ data are plotted for two subsequent approach-retraction cyles acquired at $V_{\text{sample}}=100\,\text{mV}$ [Fig.\,\ref{figS3}(a),(b)] and $V_{\text{sample}}=-100\,\text{mV}$ [Fig.\,\ref{figS3}(c),(d)] with the same tip positioned above the same graphene lattice site.
The $\Delta f(z)$ loop behavior at positve $V_{\text{sample}}$ [Fig.\,\ref{figS3}(a)] as well as its absence at negative $V_{\text{sample}}$ [Fig.\,\ref{figS3}(c)] are reproduced in the subsequent approach-retraction cycle [Fig.\,\ref{figS3}(b),(d)]\@.
Moreover, the insets to Fig.\,\ref{figS3}(a) and Fig.\,\ref{figS3}(c) reveal that STM images of the surface before and after a contact experiment are virtually identical.
These data support the idea that forming and breaking the tip--graphene covalent bond together with the partial detachment of graphene at positive $V_{\text{sample}}$ represent a reversible process.

\section{Current-voltage characteristics} 

\begin{figure}
\includegraphics[width=0.65\linewidth]{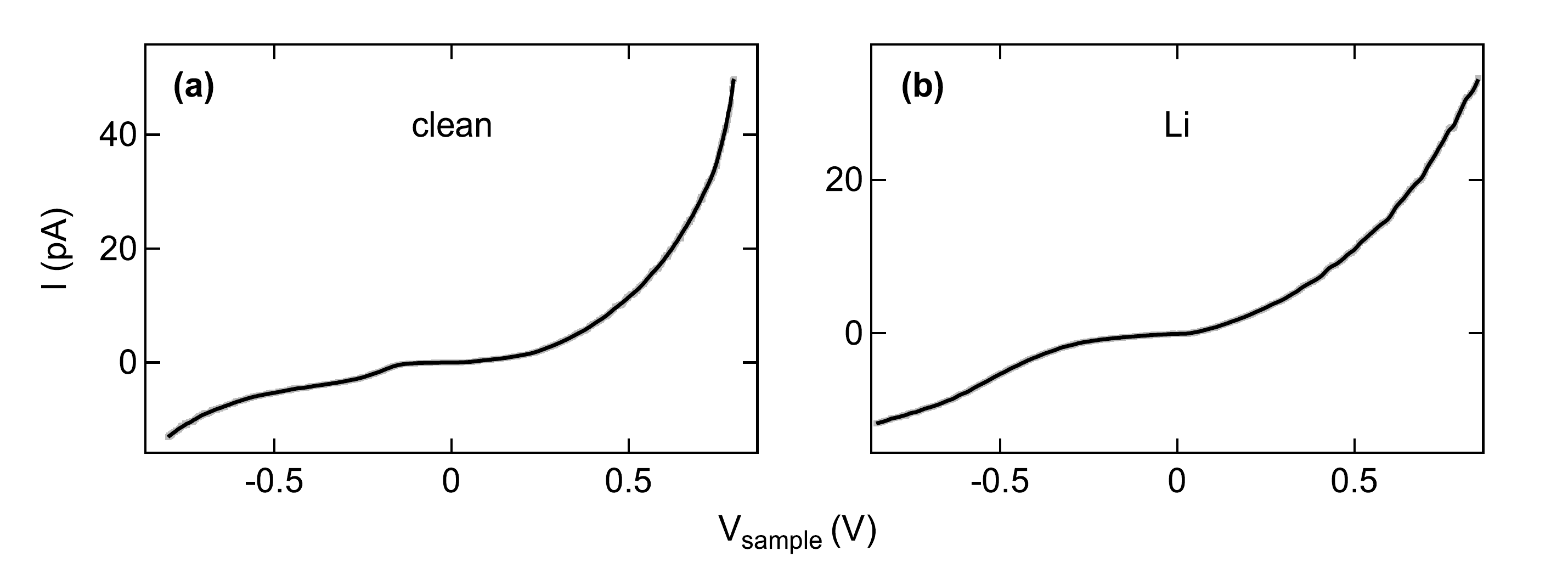}
\caption{
Current-voltage characteristics for (a) clean and (b) Li-intercalated graphene.
Feedback loop parameters: (a) $0.8\,\text{V}$, (b) $1.0\,\text{V}$, $50\,\text{pA}$\@.
}
\label{figS4}
\end{figure}

The evolution of the current as a function of $V_{\text{sample}}$ was explored in the tunneling range.
For clean graphene [Fig.\,\ref{figS4}(a)], $I(V_{\text{sample}})$ is approximately point-symmetric with respect to $V_{\text{sample}}=0\,\text{V}$ for $-0.35\,\text{V}<V_{\text{sample}}<0.35\,\text{V}$\@.
In this $V_{\text{sample}}$ interval the asymmetry of the loop width is already clearly observed [Fig.\,2(a) of the main article]\@.
For Li-intercalated graphene, $I(V_{\text{sample}})$ evolves in a similar manner [Fig.\,\ref{figS4}(b)] as observed for clean graphene.
In the $V_{\text{sample}}$ range where $I(V_{\text{sample}})$ is nearly point-symmetric with respect to $V_{\text{sample}}=0\,\text{V}$, the asymmetric loop width behavior is visible [Fig.\,2(b) of the main article]\@.
Therefore, the current across the junction can be excluded as the cause for absence and presence of  the looped $\Delta f(z)$ evolution at negative and positive $V_{\text{sample}}$, respectively.

\section{Theory}
\newcommand\sisl{\textsc{sisl}}
\newcommand\siesta{\textsc{Siesta}}
\newcommand\tsiesta{\textsc{TranSiesta}}
\newcommand\tbtrans{\textsc{TBtrans}}

The self-consistent DFT$+$NEGF method implemented in \tsiesta\ and \tbtrans\ \cite{jpcm_14_2745,prb_65_165401,cpc_212_8} and post-processing tools implemented in \sisl\ \cite{zerothi_sisl} were used.
All calculations were performed using a $300\,\text{Ry}$ mesh cut-off, single-$\zeta$ polarized basis set, and PBE$+$GGA exchange-correlation \cite{prl_77_3865}, and otherwise default parameters.
The current across the junction was simulated in a three-terminal transport calculation \cite{nl_18_5697,smallmeth_4_1900817}. 
The electric field was oriented perpendicular to the graphene plane.
In the graphene plane a $10\times 10$ $k$-mesh entered into the calculations.
Spin polarization was not considered. 
A simplified quasi-one-dimensional Au tip was placed on top of a free finite graphene sheet \cite{nanoscale_11_6153,prb_100_195417} with a Au--C distance of $225\,\text{pm}$.
For the calculation of spring constants, changes in forces with C atom displacements of $\pm 10\,\text{pm}$ and tip movements of $25\,\text{pm}$ were used.
Geometries were relaxed at zero field and $0\,\text{V}$ with a tolerance of $0.1\,\text{eV}/\text{nm}$, where in the optimization the first $4$ layers in the Au tip were relaxed as one block.
In addition, the tip-terminating Au atom as well as all C atoms around the tip within a radius of $0.75\,\text{nm}$ were subject to relaxations.
The doping level of graphene was controlled by an electrostatic bottom gate \cite{pccp_18_1025,prb_100_035415} a distance of $1.5\,\text{nm}$ underneath the graphene sheet that shifts the Dirac point to the energies observed in the experiments \cite{surfsci_699_121638,prb_93_195421}.

\subsection{Projected density of states}

\begin{figure}[h!]
\includegraphics[width=0.65\textwidth]{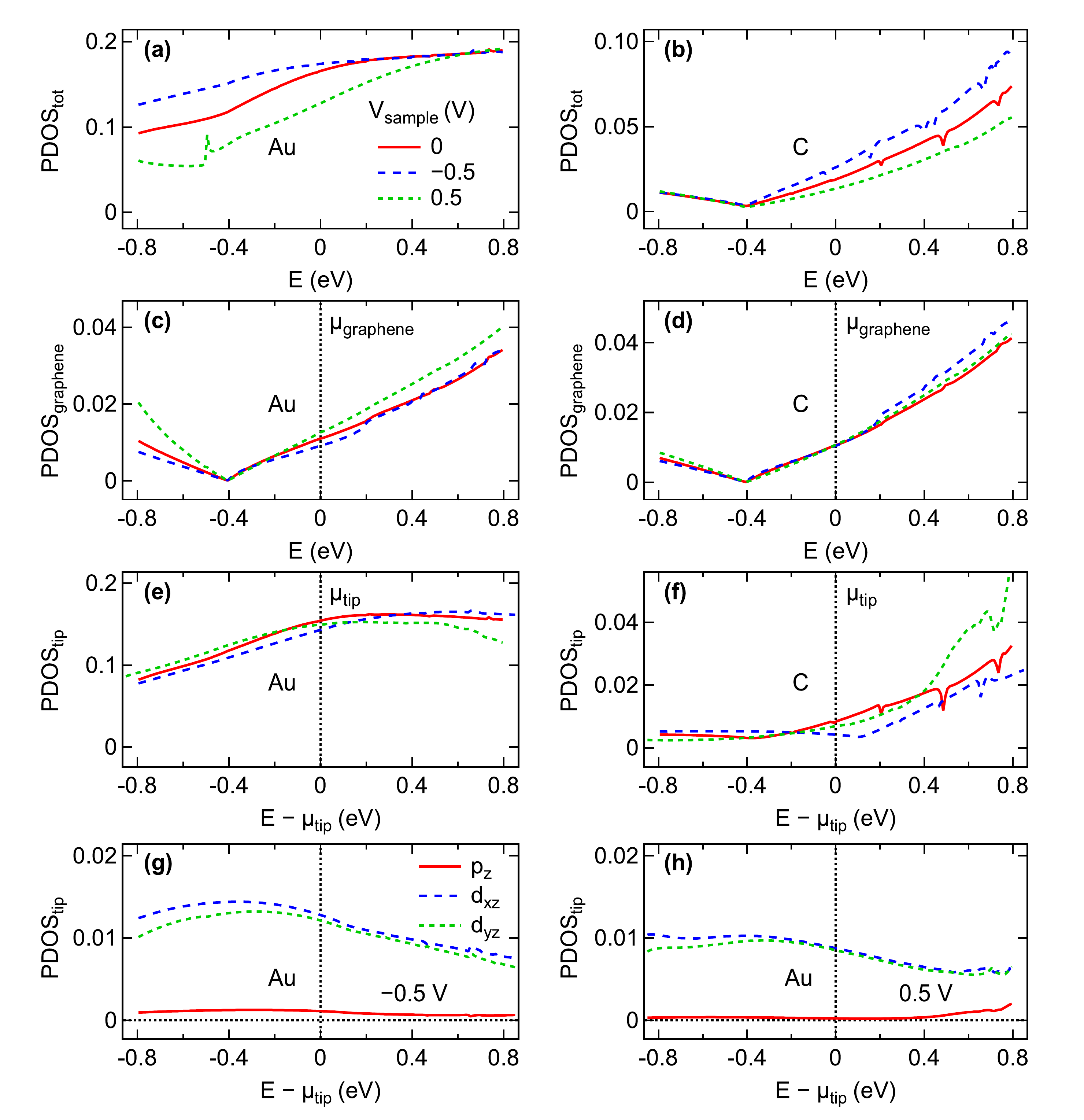}
\caption{(a), (b) Total PDOS for the Au atom at the tip apex (a) and a C atom of graphene (b) plotted for indicated $V_{\text{sample}}$.
(c)--(f) PDOS separated into scattering states originating from graphene (c),(d) and the tip (e),(f)\@.
The legend indicating $V_{\text{sample}}$ in (b)--(f) is the same as in (a)\@.
The chemical potential of graphene, $\mu_{\text{graphene}}$, is at $0\,\text{eV}$ [vertical line in (c),(d)]\@, while the chemical potential of the tip, $\mu_{\text{tip}}$ is set to $\text{e}V_{\text{sample}}$.
(g), (h) PDOS of the tip at the Au apex atom calculated for the indicated orbitals ($z$: direction of the electric field) at $V_{\text{sample}}=-0.5\,\text{V}$ (g) and $V_{\text{sample}}=0.5\,\text{V}$ (h)\@.}
\label{figS5}
\end{figure}

For finite bias voltage the total atom-projected density of states (PDOS) was found to shift, as seen for the Au atom terminating the tip apex [Fig.\,\ref{figS5}(a)] and the C atom [Fig\,\ref{figS5}(b)] to which it bonds. 
However, this shift does not reflect a major change in the electronic structure.
Rather, it shows that the scattering states originating from different electrodes (Au tip, graphene) are filled to different chemical potentials ($\mu_{\text{tip}}$, $\mu_{\text{graphene}}$)\@. 
The different filling is presented in Fig.\,\ref{figS5}(c)--(f)\@.
The PDOS was split into PDOS$_{\text{graphene}}$ [Fig.\,\ref{figS5}(c),(d)] and PDOS$_{\text{tip}}$ [Fig.\,\ref{figS5}(e),(f)], which consider the scattering states originating from, respectively, graphene and the tip, and are shifted to the corresponding chemical potential. 
These PDOS data exhibit only little change in the $V_{\text{sample}}$ range $(\mu_{\text{graphene}}-\mu_{\text{tip}})/\text{e}$ (e: elementary charge) indicating that the changes in the electronic structure due to the bias voltage is not a main factor.

Figure \ref{figS5}(g),(h) displays the PDOS on the Au tip atom along the electric-field direction, $z$,  for the orbitals $p_z$, $d_{xz}, d_{yz}$ orbitals that are most affected in the chemical bond.
They are slightly larger at negative $V_{\text{sample}}$ [Fig.\,\ref{figS5}(g)] compared to positive $V_{\text{sample}}$ [Fig.\,\ref{figS5}(h)]\@. 
This difference indicates that the charge in these orbitals changes with $V_{\text{sample}}$ in accordance with the shape of the induced charge density shown in Fig.\,4(b),(c) of the main text.

\subsection{Model of a Au--C dimer in an electric field}

\begin{figure}[h]
\includegraphics[width=0.65\linewidth]{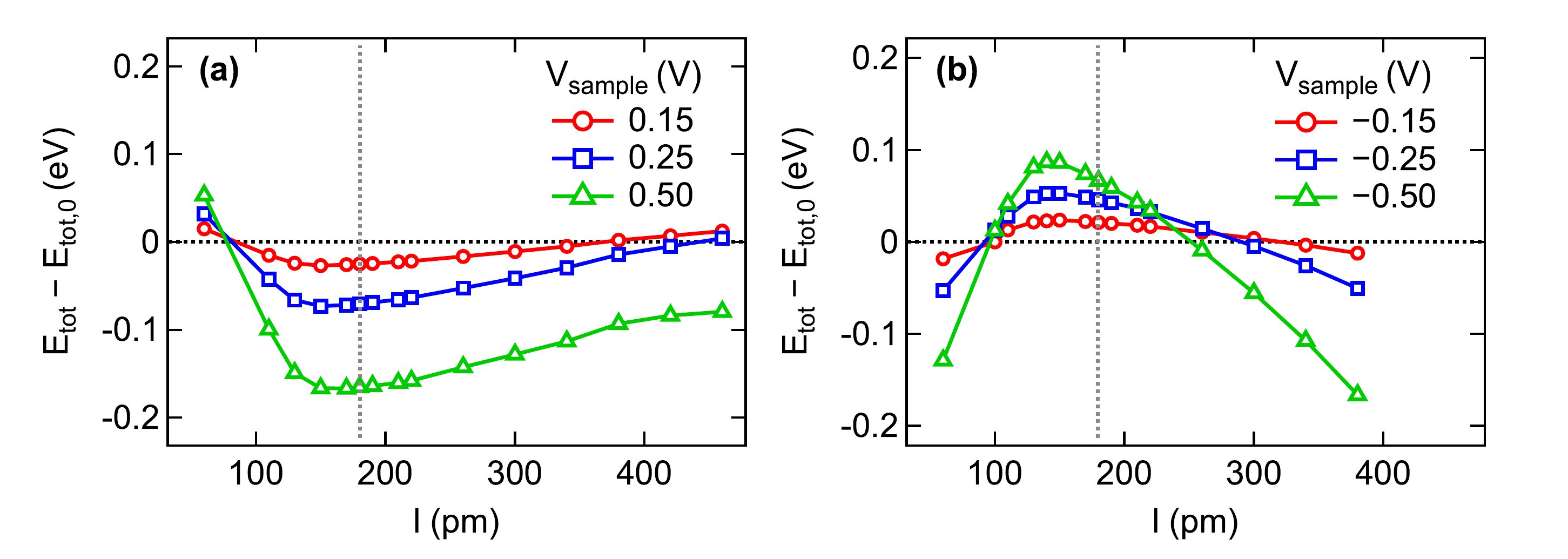}
\caption{Change in total energy $E_{\text{tot}}-E_{\text{tot},0}$ as a function of bond length $l$ for a Au--C dimer in an electrical field for indicated $V_{\text{sample}}$.
The vertical line marks the equilibrium bond length at $V_{\text{sample}}=0\,\text{V}$\@.
The field direction and strength is chosen to model positive (a) and negative (b) $V_{\text{sample}}$ of the full tip--graphene system.}
\label{figS6}
\end{figure}

Figure \ref{figS6} shows the change in total energy, $E_{\text{tot}}-E_{\text{tot},0}$ ($E_{\text{tot},0}$: total energy at $V_{\text{sample}}=0\,\text{V}$) for a Au--C dimer in an electric field as a function of the bond length $l$. 
The field is adjusted to mimic the voltage drop across the bond in the full tip--graphene calculation. 
A positive $V_{\text{sample}}$ [Fig.\,\ref{figS6}(a)] corresponds to the electric field pointing in the direction from C to Au, which forces electron transfer from Au to C and stabilizes the dipole.
This leads to an attractive force that decreases $l$ below the equilibrium bond length $l_0$ [vertical line in Fig.\,\ref{figS6}(a),(b)]\@. 
For negative $V_{\text{sample}}$ [Fig.\,\ref{figS6}(b)] repulsive forces are induced, which yield $l>l_0$.

The change in total energy with $V_{\text{sample}}$ or, equivalently, with the electric field is on the same order of magnitude as obtained for a dipole in a field. 
From Fig.\,4(a) of the main article a charge transfer of $\approx 0.1\,\text{e}$ from Au to C occurs at $V_{\text{sample}}=1\,\text{V}$\@. 
Thus, a total energy change on the order of $0.1\,\text{eV}$ can be estimated, which is also seen in Fig.\,\ref{figS6}(a)\@. 
Because of the charge transfer and the electric dipole increasing linearly with $V_{\text{sample}}$, a beyond-linear increase of total energy changes with $V_{\text{sample}}$ may be expected and is seen especially in Fig.\,\ref{figS6}(a)\@.

\bibliographystyle{apsrev4-2}
%